# Explainable artificial intelligence in breast cancer detection and risk prediction: A systematic scoping review

Amirehsan Ghasemi[1,2]  | Soheil Hashtarkhani[1]  | David L. Schwartz[3]  | Arash Shaban-Nejad[1,2]

[1]Department of Pediatrics, Center for Biomedical Informatics, College of Medicine, University of Tennessee Health Science Center, Memphis, Tennessee, USA

[2]The Bredesen Center for Interdisciplinary Research and Graduate Education, University of Tennessee, Knoxville, Tennessee, USA

[3]Department of Radiation Oncology, College of Medicine, University of Tennessee Health Science Center, Memphis, Tennessee, USA

**Correspondence**
Arash Shaban-Nejad, Department of Pediatrics, Center for Biomedical Informatics, College of Medicine, University of Tennessee Health Science Center, Memphis, TN 38103, USA.
Email: ashabann@uthsc.edu

**Funding information**
None

## Abstract

With the advances in artificial intelligence (AI), data-driven algorithms are becoming increasingly popular in the medical domain. However, due to the nonlinear and complex behavior of many of these algorithms, decision-making by such algorithms is not trustworthy for clinicians and is considered a black-box process. Hence, the scientific community has introduced explainable artificial intelligence (XAI) to remedy the problem. This systematic scoping review investigates the application of XAI in breast cancer detection and risk prediction. We conducted a comprehensive search on Scopus, IEEE Explore, PubMed, and Google Scholar (first 50 citations) using a systematic search strategy. The search spanned from January 2017 to July 2023, focusing on peer-reviewed studies implementing XAI methods in breast cancer datasets. Thirty studies met our inclusion criteria and were included in the analysis. The results revealed that SHapley Additive exPlanations (SHAP) is the top model-agnostic XAI technique in breast cancer research in terms of usage, explaining the model prediction results, diagnosis and classification of biomarkers, and prognosis and survival analysis. Additionally, the SHAP model primarily explained tree-based ensemble machine learning models. The most common reason is that SHAP is model agnostic, which makes it both popular and useful for explaining any model prediction. Additionally, it is relatively easy to implement effectively and completely suits performant models, such as tree-based models. Explainable AI improves the transparency, interpretability, fairness, and trustworthiness of AI-enabled health systems and medical devices and, ultimately, the quality of care and outcomes.

# 1 | INTRODUCTION

Breast cancer (BC) is one of the most common cancers with high morbidity and mortality globally. Early detection and treatment significantly increase the chances of survival [1]. With the growing interest in artificial intelligence (AI), computer-aided diagnosis (CAD) based on AI has become a valuable tool for the detection, classification, and diagnosis of cancer biomarkers and morphological features.

Compared to rule-based systems [2] that require human intervention in decision-making, AI models can learn from medical data and generate new patterns by themselves. However, AI systems are susceptible to several biases [3], mostly stemming from low-quality datasets, faulty algorithms, and human cognitive biases that may lead to inaccurate decisions, predictions, or inferences [4]. Additionally, many AI systems have raised concerns among clinicians about accountability, fairness of AI algorithms, and lack of transparency [5], a critical factor for high-stakes domains such as healthcare, where a minor error in decision-making can lead to irreparable consequences [6].

Despite significant progress over the last years in terms of fine-tuning [7] and optimizing [8–11] AI algorithms to tackle supervised and unsupervised tasks, a considerable number of these algorithms remain enigmatic, classified as Black-box and are yet to be demystified. To this end, the scientific community has started investigating techniques and methods to make AI algorithms more understandable, explainable, and interpretable. In recent years, explainable artificial intelligence (XAI), coined by DARPA [12], has emerged as a notable and noteworthy topic of discussion in the AI community. The rationale behind XAI lies in the assumption that such techniques establish rules for more trustworthy AI systems by making them more transparent, understandable, interpretable, safe, and reliable while making a decision or recommending an action [13, 14].

Typically, AI models are evaluated based on their prediction errors [15] without providing enough transparency to the end users throughout this process. As Figure 1 shows, XAI methods are formulated to be applied to the result and provide transparency; however, the human-in-the-loop (HITL) concept must also be applied to achieve trustworthiness. XAI is basically employing human–agent interaction methods to utilize human knowledge and intuition to comprehend the rationale behind the results it gains [16]. As Figure 1 illustrates, a result generated by the AI model passes through a suitable XAI method. Then, the human agent benefits from the transparency created by the XAI to validate, confirm, or enhance the predictions [17]. It should be noted that the human agent's decision is based on the collaboration between clinicians and XAI experts. If the results are not correct or satisfying, the XAI module investigates the model, data, or both and reruns the outcomes to the AI system throughout an iterative process until a consensus over the results is reached.

Most of the review articles on XAI models investigate the application of XAI in general healthcare [18–20]. Although there have been reviews of the subject on other types of cancer [21], to the best of our knowledge, our contribution to reviewing existing XAI technologies in breast cancer screening, risk detection, and prediction is distinctive both in terms of scope and breadth. This paper provides a comprehensive summary of published studies and then elaborates on the background and concepts associated with the XAI methods used in these studies. Finally, we highlight the most popular XAI methods and explain the rationale behind their popularity.

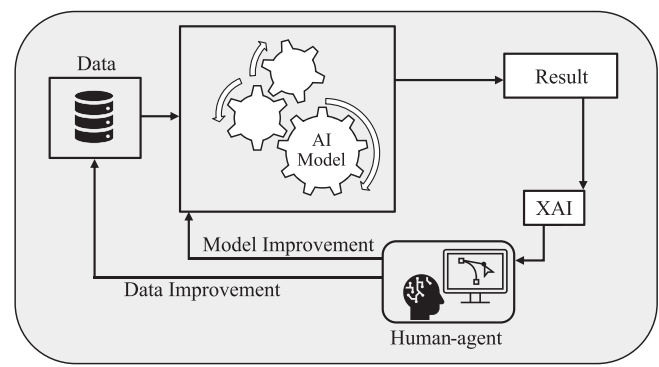

**FIGURE 1** Explainable artificial intelligence (XAI) as a human-agent problem-solving method.

The remainder of this paper is organized as follows. Section 2 covers the background, including the introduction to AI models, the concept of accuracy-explainability trade-off, and the classification of XAI methods. Section 3 describes the research method. Section 4 provides tabulated results of implemented XAI methods in breast cancer research. Section 5 discusses the details and elaborates on the utilized methods. Finally, Section 6 concludes our survey by highlighting XAI's achievements, strengths, and limitations and discussing future research opportunities.

## 2 | BACKGROUND

### 2.1 | Overview of AI models

AI models employ data-driven algorithms to reach decisions or identify explanatory patterns. Machine learning (ML) algorithms fall into three types: regression and classification, which are supervised, and clustering, which is unsupervised. If the output is a continuous variable, we deal with regression, but when it is discrete labels or categories, then we use classification [22]. Clustering algorithms identify and group similar data points based on their characteristics. The most popular ML models used in breast cancer studies are listed in Table 1.

Deep learning (DL) is a subcategory of ML that may be supervised or unsupervised. Unlike traditional machine learning, DL models require much less manual human intervention since they automate the feature extraction, saving time and resources. DL models are capable of addressing model accuracy and performance for unstructured data such as speech, images, videos, or texts, whereas classical ML models do not function effectively.

In BC datasets, we usually deal with high-dimensional, multimodal [30] structured and unstructured data, which are often big, noisy, and sparse, making them challenging to analyze. Thanks to neural networks' universal approximation [31, 32] and the advantage of auto-differentiation [33], deep learning models can be applied to many of these problems. DL models can learn the complex nonlinear relationships between the features and target variables, making them viable data-driven models that enable new discoveries in breast cancer classification and detection. Frequently used DL models and their variants in BC studies are listed in Table 2.

**TABLE 1** List of popular machine learning (ML) models.

| ML model | Acronym | Type of learning | Type of problem |
| --- | --- | --- | --- |
| Linear regression | N/A | Supervised | Regression |
| Logistic regression | LR | Supervised | Classification |
| Decision trees | DT | Supervised | Regression, classification |
| K-means | N/A | Unsupervised | Clustering |
| Naïve Bayes | NB | Supervised | Classification |
| Support vector machines | SVM | Supervised | Regression, classification |
| K-nearest neighbors | KNN | Supervised | Regression, classification |
| Ensemble learning models[a] | | | |
| Extremely randomized trees [23] | Extra-trees (ET) | Supervised | Regression, classification |
| Random forests [24] | RF | | |
| Gradient boosting machines | GBM | | |
| eXtreme gradient boosting [25] | XGBoost | | |
| Light gradient boosting machine [26] | LightGBM | | |
| Gradient boosted decision trees | GBDT | | |
| Adaptive boosting [27] | AdaBoost | | |
| Category boosting [28] | CatBoost | | |

[a]Ensemble learning is a meta-learning approach that combines multiple models to make a decision, typically in supervised ML tasks [29].

**TABLE 2** List of popular deep learning (DL) models.

| DL model | Acronym | Variants | Acronym |
|---|---|---|---|
| Convolutional neural network [34] | CNN | Visual geometry group [35] | VGG (VGG-16, VGG-19) |
| | | AlexNet [36] | N/A |
| | | Xception [37] | N/A |
| | | GoogLeNet [38] | N/A |
| | | GoogLeNet inception V3 [39] | Inception V3 |
| | | GoogLeNet inception V4 [40] | Inception V4 |
| | | Residual networks [41] | ResNet |
| | | ResNeXt [42] | N/A |
| | | ResNet (Split attention networks) [43] | ResNeSt |
| | | U-Net [44] | N/A |
| | | Graph convolutional network [45] | GCN |
| | | Dense convolutional network [46] | DenseNet |
| | | EfficientNet [47] | N/A |
| | | MobileNet [48–50] | N/A |
| | | ShufflieNet [51] | N/A |
| | | SqueezeNet [52] | N/A |
| Recurrent neural network | RNN | Long short-term memory [53] | LSTM |
| | | Bidirectional LSTM [54] | BiLSTM |
| | | Gated recurrent unit [55] | GRU |

## 2.2 | Accuracy-explainability trade-off

As Figure 2 illustrates, the accuracy-explainability trade-off refers to the balance between the accuracy of an AI model and its explainability [56]. The goal of any AI model is to generate highly accurate results. From the XAI perspective, the models must be explainable. However, achieving both accuracy and explainability is far from trivial. Regarding explainability, AI models can be black-box, white-box, or gray-box [12, 57], as depicted in Figure 2.

White-box models are intrinsically transparent and explainable [58]. However, they are limited to learning only linear associations between input features and the target variable. Although white-box models may not achieve high accuracy levels, they offer human-understandable explanations. In contrast, black-box models are nontransparent by nature [59]. While these models may have outstanding performance, they suffer from a lack of explainability. Gray-box models strike a balance between accuracy and explainability. Generally, any data-driven learning algorithm, including black- and white-box models, is considered a gray box [57, 60]. For a gray-box model, connections from input data to model output can be explained despite not being fully transparent [60].

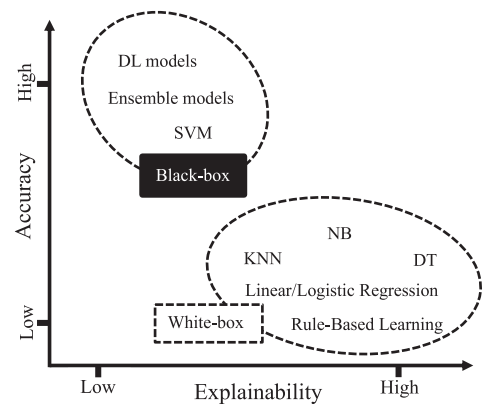

**FIGURE 2** Trade-off between model accuracy and explainability.

## 2.3 | Classification of XAI methods

The transparency of AI systems can be addressed from different perspectives. The results of XAI methods can be presented in various ways, including numerical, rules, textual, visual, or a combination of these [61]. As Figure 3 depicts, three critical factors for categorizing XAI methods of explanation exist: scope, stage, and type.

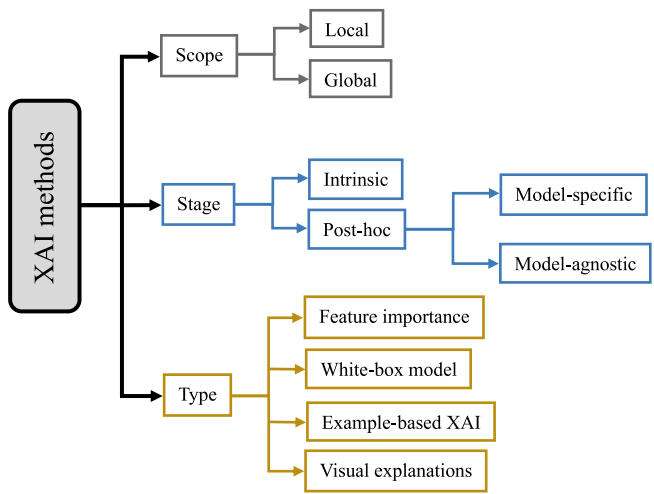

**FIGURE 3** Classification of explainable artificial intelligence (XAI) methods.

The **scope** of explainability is either local or global. The local method aims to shed light on an AI model for a specific input [62, 63] and explains why a particular decision was made by highlighting the input feature influencing the model's output. However, this approach cannot find a general relationship between input features and outputs [64]. The global method provides a broader understanding by analyzing the model's overall structure and general patterns across the entire data set or a larger subset, helping users understand its biases, limitations, and general decision-making patterns [62, 63].

Intrinsic and post hoc [65] refer to the **stage** of explanation. The intrinsic approach refers to using white-box models which are interpretable by nature. The post hoc approach relates to explainable methods that "explain the model predictions after the training and inference processes" [62, 64]. Generally, the post hoc approach, compared to intrinsic models (white-box models) [65], is more accurate since post hoc explainable methods must be applied to black-box models' prediction, and black-box models tend to perform better in results. Although some of these methods, such as rule extraction [66] and tree extraction [67], can turn black-box models into white-box, there is a complexity–accuracy trade-off [57, 64]. Additionally, post hoc methods are either model-specific or model-agnostic. Model-specific methods are designed to explain specific black-box models by investigating their internal factors and interactions [64]. For example, many techniques are developed to analyze DL models, which attempt to find the contribution of artificial neurons on their final decisions through backpropagation (backprop) error [68–71]. Model-agnostic methods provide explanations independent of a specific AI model. Some common post hoc XAI methods are tabulated in Appendix A.

XAI methods can be classified based on the **type** of explanations they offer [64]. As Figure 3 demonstrates, there are four types of explanations: feature importance, white-box model, example-based XAI, and visual explanations [64]. For the first type, XAI methods create numbers/values for the input features to express the feature's importance. For the second type, XAI methods "create a white-box model that mimics the original black-box model and is inherently explainable" [64]. The example-based type, also known as data point [65], uses samples from the training datasets to explain the model's action. For the last type, XAI methods offer a type of explainability based on purely visual explanations [64].

## 3 | RESEARCH METHOD

This systematic review is carried out using the preferred reporting items on systematic reviews and meta-analysis (PRISMA) [72] guideline in three steps as follows:

**Step 1: Identifying studies**—As mentioned in the introduction, our focus for this paper was on studies that examined existing XAI methods in breast cancer research. We conducted a comprehensive search utilizing some of the most popular and trusted citation platforms [73], including Scopus, IEEE Xplore, PubMed, and Google Scholar (first 50 citations) from January 2017 to July 2023 using the combination of keywords and MeSH terms described in Table 3. A total of 193 studies were included in this step.

**Step 2: Selecting the studies**—In Step 2, we selected articles for inclusion based on specific criteria: they had to be original studies published in peer-reviewed English-language journals, utilizing at least one XAI methodology within the context of breast cancer. Two reviewers (Amirehsan Ghasemi and Soheil Hashtarkhani), screened citations by title and abstract, excluding various types of irrelevant papers, such as different review papers ($n = 16$), those discussing XAI but unrelated to breast cancer ($n = 18$), those discussing breast cancer but unrelated to XAI ($n = 20$), preprints awaiting peer review ($n = 7$), conference papers ($n = 34$), duplicate titles ($n = 37$), and nonresearch materials like books, dissertations, editorials, and technical notes ($n = 12$), resulting in the exclusion of 144 studies. Subsequently, 49 articles underwent full-text scrutiny, with inaccessible or irrelevant articles being excluded. This left us with 30 articles that met the inclusion criteria for our comprehensive review. Figure 4 illustrates a summary of our search strategy and steps.

**TABLE 3** Explored databases and the results.

| | Boolean search strings | Number of search results |
|---|---|---|
| **Database** | | |
| Scopus | TITLE-ABS-KEY ("Explainable Artificial Intelligence" OR "Explainable AI" OR "XAI" OR "Explainable Machine Learning" OR "Interpretable Machine Learning" OR "Interpretable AI") AND TITLE-ABS-KEY ("Breast Cancer") AND PUBYEAR > 2016 | $n = 104$ |
| IEEE Xplore | ("Abstract":"Explainable Artificial Intelligence" OR "Abstract":"Explainable AI" OR "Abstract":"XAI" OR "Abstract":"Explainable Machine Learning" OR "Abstract":"Interpretable Machine Learning" OR "Abstract":"Interpretable AI") AND ("Abstract":"Breast Cancer") | $n = 9$ |
| PubMed | ("Breast Neoplasms"[Mesh] OR "breast cancer") AND ("XAI" OR "Interpretable Machine Learning" OR "Explainable Artificial Intelligence" OR "Explainable AI") string from 2017 | $n = 30$ |
| **Web search engine** | | |
| Google Scholar | ("Explainable Artificial Intelligence" OR "XAI" OR "Explainable AI") AND "Breast Cancer" | $n = 50$ |

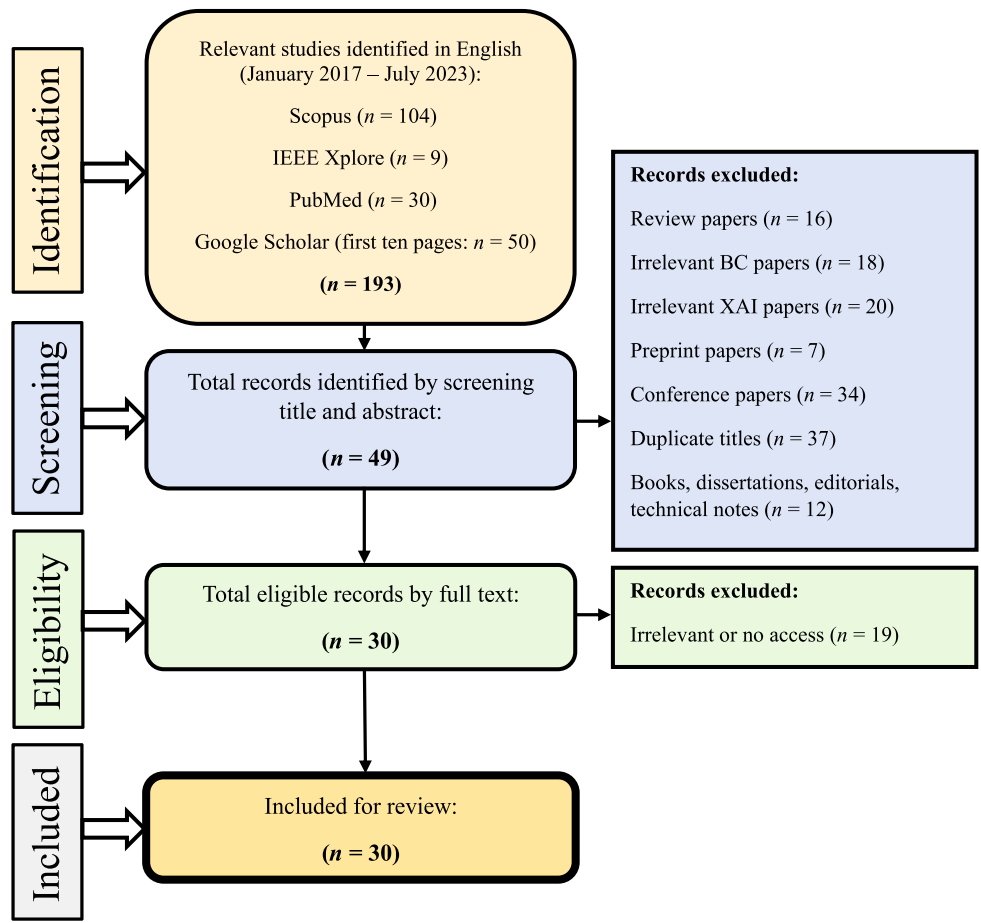

**FIGURE 4** Preferred reporting items on systematic reviews and meta-analysis guideline for article selection.

**Step 3: Data extraction and summarization**—A data extraction form was developed in Google Sheets, consisting of eight variables, including authors, year, the aim of the study (objective), data set(s), data type, important features, type of AI (ML or DL), and the explained model. Two reviewers (Amirehsan Ghasemi and Soheil Hashtarkhani) extracted data from all included articles, and any disagreement was resolved by consensus.

# 4 | RESULTS

Almost 30 studies have been identified in the literature utilizing XAI methods in breast cancer settings. XAI methods used in these studies include SHAP, LIME, CAM, Grad-CAM, Grad-CAM++, and LRP. The number of studies and explained AI models for each method are shown in Figure 5. Tables 4–9 provide tabular representations of the included studies. A detailed description of the results is provided in the discussion section (Section 5).

# 5 | DISCUSSION

## 5.1 | Post hoc XAI methods: Model-agnostic

### 5.1.1 | SHapley Additive exPlanations (SHAP)

SHAP [116] offer local and global explanations based on the Shapley value [117], a solution concept used in cooperative game theory. In SHAP, the input features of an observation act as players in a game, and the prediction serves as the reward. SHAP computes the average marginal contribution of each player to the reward [64, 65] and ensures that the distribution of reward among players is fair [18, 118]. In BC studies, SHAP can potentially find the contribution of biomarkers (players → important features in Table 4) to the prediction (reward objective in Table 4).

As provided in Table 4, most studies (12/13) implemented ensemble ML learning as the predictors. Only in one study (1/13) [89] did authors first utilize LR-based to discriminate between upregulated and regular expression of HER2 protein, then pathologists' diagnoses (IHC) in conjunction with fluorescent in situ hybridization (IHC + FISH) were used as the training outputs. In Chakraborthy et al. [74], SHAP showed that "by boosting the B cell and CD8$^+$ T cell fractions or B cell and NK T cell fractions in the tumor microenvironment (TME) to levels above their inflection points, the survival rate of BC patients could increase by up to 18%." In Rezazadeh et al. [76], texture analysis of the ultrasound images based on the gray-level co-occurrence matrix (GLCM) predicted the likelihood of malignancy of breast tumors. SHAP was used to find the most critical features: GLCM correlation and GLCM energy within different pixel distances along the 90° direction.

In summary, SHAP emerged as the most frequently used XAI method in the BC studies (13/30). Notably, no DL models were used in conjunction with SHAP. Instead, tree-based ensemble learning ML models, specifically XGBoost (9/13 studies), were the most widely used models. This can be attributed to the high-speed SHAP algorithm, which is well-suited for tree-based models such as XGBoost, Catboost, GBM, AdaBoost, and so on [18].

### 5.1.2 | Local interpretable model agnostic explanations (LIME)

LIME [119] provides a local explanation using a surrogate model. As outlined in Table 5, LIME is utilized in 5 out of 30 studies to explain the model prediction by highlighting the contribution of the most important features. LIME creates a linear local surrogate mode that is intrinsically interpretable around a sample (data point) and improves transparency by producing feature importance values. The surrogate model in LIME modifies some parts of the given features and generates perturbed instances to understand how the output changes. The perturbation depends on the nature of the input sample. For instance, one method to perturb an image is by replacing certain parts with gray color [120]. In Kaplun et al. [90], to explain the image classification, LIME puts a mask of yellow pixels to highlight the important image segments the model focuses on to make the decision.

In Adnan et al. [93], the authors have implemented SHAP in conjunction with LIME to explain that a small number of highly compact and biological gene cluster features resulted in similar or better performance than classifiers built with many more individual genes. With training on smaller gene clusters, LIME proved that the classifiers have better AUC than the original classifiers except in RF and rSVM. In Saarela and Jauhiainen [92], the authors used linear and nonlinear ML classifiers with

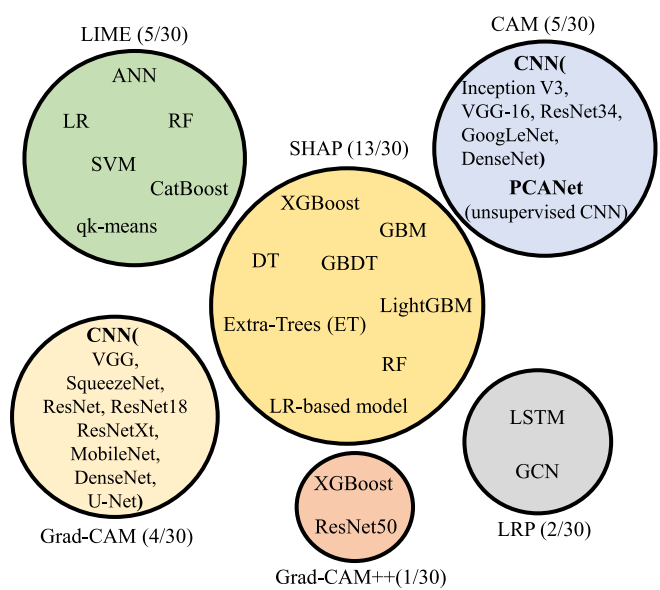

**FIGURE 5** Explainable artificial intelligence (XAI) methods and explained models used in the literature.

**TABLE 4** List of studies that used SHapley Additive exPlanations.

| Authors | Year | Objective | Data set(s) | Data type | Important features | ML/DL | Explained model |
|---|---|---|---|---|---|---|---|
| Chakraborty et al. [74] | 2021 | Investigate the relationship between immune cell composition in the tumor microenvironment (TME) and the ≥5-year survival rates of breast cancer patients | Patient clinical information for TCGA breast invasive carcinoma data from two projects on the cbioPortal | Clinical data | B cells, CD8+ T cells, NK T cells, M0 macrophages | ML | XGBoost |
| Moncada-Torres et al. [75] | 2021 | The Cox Proportional Hazards (CPH) (identifying the prognostic factors that have an impact on patients' recurrence or survival) | Netherlands Cancer Registry (NCR) 36,658 non-metastatic breast cancer patients | Text | Age, pts, ptmm | ML | XGBoost |
| Rezazadeh et al. [76] | 2022 | Breast cancer diagnosis based on the gray-level co-occurrence matrix (GLCM) | Data set of breast ultrasound images [77] | Ultrasound | GLCM texture features | ML | Decision tree Ensemble model (DT, GBDT, LightGBM) |
| Nahid et al. [78] | 2022 | Classify BC patients and non-BC patients through regular examination of a few health-related issues such as the level of Glucose, Insulin, HOMA, Leptin, etc | Data set by the University Hospital Centre of Coimbra. 116 participants (64 BC, 52 non-BC) | Text (blood test data) | Glucose | ML | GBM |
| Yu et al. [79] | 2022 | Clarify the radiation dose-volume effect of radiation therapy to avoid radiation-induced lymphopenia | 589 patients with breast cancer who underwent radiation therapy at the University of Hong Kong-Shenzhen Hospital | Clinical data | Baseline lymphocyte counts protect against while the baseline hemoglobin level impacts the event of radiation-induced lymphopenia | ML | XGBoost |
| Meshoul et al. [80] | 2022 | Improve the multiclassification performance of ML models for BC cancer subtyping for high dimensional datasets with a minimal number of instances | The Cancer Genome Atlas (TCGA) | Omics data | DNA, RNA, CNV | ML | Extra-Trees (ET) |
| Kumar et al. [81] | 2023 | Identify potential diagnostic biomarkers for BC | NCBI-GEO Database: two datasets were identified (GSE27562, GSE47862) (252 breast cancer patients and 194 healthy women) | Peripheral blood mononuclear cells (PBMC) (Genomic data) | SVIP, BEND3, MDGA2, LEFI-AS1, PRM1, TEX14, MZB1, TMIGD2, KIT, FKBP7 | ML | XGBoost |
| Silva-Aravena et al. [82] | 2023 | Developing a clinical decision support methodology that performs early detection of BC and interprets the variables and how they affect patients' health | Public data on women from Indonesia [83] (400 anonymous patient cases, 200 of them with BC) | Text | High-fat diet, breastfeeding | ML | XGBoost |

**TABLE 4** (Continued)

| Authors | Year | Objective | Data set(s) | Data type | Important features | ML/DL | Explained model |
|---|---|---|---|---|---|---|---|
| Massafra et al. [84] | 2023 | Predict 5-year and 10-year breast cancer invasive disease events (IDEs) | (486 breast cancer patients) Breast and clinic research center IRCCS Istituto Tumori "Giovanni Paolo II" in Bari (Italy) | Clinical data | (5 years) Age, tumor diameter, surgery type, multiplicity. (10 years) therapy-related features: hormone, chemotherapy schemes, lymphovascular invasion | ML | XGBoost |
| Vrdoljak et al. [85] | 2023 | Assessing metastatic lymph node status in BC patients eligible for neoadjuvant systemic therapy (NST) | Data collected from all Croatian hospitals (total study population (8381), NST-criteria group (719)) | Text | NST group: (tumor size, ER, PR, HER2), Total population: (tumor size, Ki-67, tumor grade) | ML | NST-criteria group (RF), Total study population → (XGBoost) |
| Uddin et al. [86] | 2023 | Investigate an ML model to forecast the development of BC more promptly | Breast Cancer Wisconsin (Diagnostic) [87] | Text | **Results for LightGBM:** perimeter_worst, concave points_mean, concave points_worst | ML | LightGBM, XGBoost, GBM |
| Zhao et al. [88] | 2023 | Predicting distant metastasis in male breast cancer (MBC) patients | 2241 MBC patients from the SEER database between 2010 and 2015, and 110 MBC patients from a hospital between 2010 and 2020 | Clinical and pathological TNM staging information data | T stage, age, N stage | ML | XGBoost |
| Cordova et al. [89] | 2023 | Classifying the epidermal growth factor 2 (HER2) photomicrographs to determine criteria that improve the value of immunohistochemical (IHC) analysis | 393 histological slides of IHC-stained breast cancer tissues from 2019 were randomly collected for the lab technician team of Carlos Van Buren Hospital Pathology Service (Valparaíso, Chile) | Microscopy images | **Results for IHC + FISH:** COUNT, MGV, M. SIZE, % AREA. **Results for IHC:** MGV, COUNT, M. SIZE, %AREA | ML | LR-based to discriminate between upregulated and normal expression of HER2 protein |

**TABLE 5** List of studies that used local interpretable model agnostic explanations.

| Authors | Year | Objective | Data set(s) | Data type | Important features | Machine learning (ML)/ Deep learning (DL) | Explained model |
|---|---|---|---|---|---|---|---|
| Kaplun et al. [90] | 2021 | Extract complex features from cancer cell images and classify malignant and benign cancer cell images | BreakHis [91] | Microscopic images | Yellow highlighted segments in the image | DL | ANN (2-layer feed forward neural network) |
| Saarela et al. [92] | 2021 | Comparing different feature importance measurements using linear (LR) and nonlinear (RF) classification ML models | Breast Cancer Wisconsin (Diagnostic) [87] | Text | L1-LR → all except one (compactness 3) RF → nine features were significant | ML | L1 regularized LR, RF |
| Adnan et al. [93] | 2022 | Proposing a model in BC metastasis prediction that can provide personalized interpretations using a very small number of biologically interpretable features | Amsterdam Classification Evaluation Suite (ACES) [94] (composed of 1616 patients, among which 455 is metastatic) | Genomic data | N/A | M/DL | RF, LR, lSVM, rSVM, ANN |
| Maouche et al. [95] | 2023 | Propose an explainable approach for predicting BC distant metastasis that quantifies the impact of patient and treatment characteristics | Public data set composed of 716 Moroccan women diagnosed with breast cancer [96] | Clinicopathological data | The characteristics have different impacts ranging from high, moderate, and low | ML | Cost-sensitive CatBoost |
| Deshmukh et al. [97] | 2023 | Improve the qk-means clustering algorithm using LIME to explain the predictions | The breast cancer data set has 600 attributes or patient records and 7 features | Text | A tabular explainer explains the positively and negatively correlated features | ML | qk-means (hybrid classical-quantum clustering approach) |

**TABLE 6** List of studies that used class activation map.

| Authors | Year | Objective | Data set(s) | Data type | Machine learning (ML)/Deep learning (DL) | Explained model |
|---|---|---|---|---|---|---|
| Qi et al. [98] | 2019 | Improving the efficiency and reliability of BC screening and guiding pathological examination by automating ultrasonography image diagnosis | Department of Galactophore Surgery and Department of Oncology of West China Hospital, Sichuan University (Over 8000 images from 2047 patients from October 2014 to August 2017) | Ultrasound | DL | Convolutional neural network (CNN) |
| Zhou et al. [99] | 2019 | Predicting clinically negative axillary lymph node metastasis from images in patients with primary breast cancer | Tongji Hospital (974 images (2016 to 2018)), independent test set (Hubei Cancer Hospital (81 imaging (2018 to 2019)) | Ultrasound | DL | Inception V3 |
| Huang et al. [100] | 2020 | Propose unsupervised DL learning model for medical image classification | CBIS-DDSM: Breast Cancer Image Data set | X-ray | DL | Modified PCANet (An unsupervised CNN model), DenseNet |
| Xi et al. [101] | 2020 | Proposing a DL-based approach for abnormality detection in medical images | (1) Mammographic Image Analysis Society (MIAS), (2) Digital Database for Screening Mammography (DDSM) | X-ray | DL | CNN |
| Kim et al. [102] | 2021 | Developing a weakly-supervised CNN algorithm to diagnose breast cancer without using image annotation | 1400 US images for breast masses of 971 patients from two institutions | Ultrasound | DL | VGG-16, ResNet34, GoogLeNet |

**TABLE 7**  List of studies that used Grad-class activation map.

| Authors | Year | Objective | Data set(s) | Machine learning (ML)/Deep learning (DL) | Explained model |
|---|---|---|---|---|---|
| Adoui et al. [103] | 2020 | Predicting the breast cancer response to Neoadjuvant chemotherapy (NAC) based on multiple MRI inputs | Institute of Radiology in Brussels (A cohort of 723 axial slices extracted from 42 breast cancer patients who underwent NAC therapy) | MRI | DL | Based on convolutional neural network (CNN) |
| Hussain et al. [104] | 2022 | Developing DL multiclass shape-based classification framework for the tomosynthesis of breast lesion images | Based on the previous study [105] | Digital breast tomosynthesis (DBT) | DL | VGG, ResNet, ResNeXt, DenseNet, SqueezeNet, MobileNet-v2 |
| Agbley et al. [106] | 2023 | Breast tumor detection and classification using different magnification factors on the Internet of Medical Things (IoMT) | BreakHis [91] | Microscopic images | DL | ResNet-18, Federated Learning (FL) to preserve the privacy of patient data |
| Gerbasi et al. [107] | 2023 | Proposing a fully automated and visually explained model to analyze raw mammograms with microcalcifications | INbreast data set [108] (train and test), CBIS-DDSM [109] (used to implement the classification algorithm) | Scanned film Mammography | DL | U-Net, ResNet18 |

LIME to understand how they differ in explaining features' importance. It was found that the nonlinear model (RF) offered better explainability as it focused on fewer features (9) compared to the nonlinear model (all except one feature). Deshmukh et al. [97] used LIME to quantify the impact of patient and treatment characteristics on BC distant metastasis. It reached the results of different impacts ranging from high impacts, such as the nonuse of adjuvant chemotherapy, to moderate impact of carcinoma with medullary features cancer type, to a low impact of oral contraception use.

As a model-agnostic method, LIME was used to explain various ML models, including RF, SVM, ensemble learning, and a shallow DL learning model, as detailed in Kaplun et al. [90]. LIME only offers a local interpretation, and compared to SHAP, when a large volume of predictions needs to be explained, it has a higher speed and can be a more excellent alternative. In summary of the model-agnostic methods used in studies (18/30), SHAP was preferred over LIME to shed light on the most important features. This is because SHAP is relatively easy to implement and provides both local and global explanations, and compared to LIME, it has a higher speed on the global-level explanation for high-performance ensemble ML models.

## 5.2 | Post hoc XAI methods: Model-specific

Several XAI methods are specifically designed for different DL architectures focusing on the feature importance type of explanation. Most of these methods are propagation-based and enjoy the availability of gradients computed during the training.

### 5.2.1 | Class activation map (CAM)

CAM [121] is a local backpropagation-based method that uses a global average pooling (GAP) layer after the last convolutional layer, followed by the classification layer to identify the most discriminative regions of an image in the convolutional neural network (CNN) [34]. This technique combines a linear weighted sum of the feature maps to gain a heatmap that highlights class-specific regions of the image. CAM is limited to existing networks that have the described architecture.

As listed in Table 6, (5/30) studies have used CAM to determine how accurately the CNN model localized the breast masses. Qi et al. [98] proposed two CNN-based networks, the Mt-Net and the Sn-Net, to identify malignant tumors and recognize solid nodules step-by-step. To enable the two networks to collaborate effectively, CAM was introduced as an enhancement mechanism to improve the accuracy and sensitivity of the classification results for both networks.

**TABLE 8**  List of studies that used Grad-class activation map++.

| Authors | Year | Objective | Data set(s) | Data type | Machine learning (ML)/Deep learning (DL) | Explained model |
|---|---|---|---|---|---|---|
| To et al. [110] | 2023 | Improving classification performance and effectively identifying cancerous regions in DUV whole-slide images (WSI) | Medical College of Wisconsin (MCW) tissue bank [111] (60 samples, 24 normal/benign and 36 malignant) | DUV-WSI image | ML/DL | ResNet50, XGBoost |

**TABLE 9**  List of explainable artificial intelligence studies that used layerwise relevance propagation.

| Authors | Year | Objective | Data set(s) | Data type | Machine learning (ML)/Deep learning (DL) | Explained model |
|---|---|---|---|---|---|---|
| Grisci et al. [112] | 2021 | Propose relevance aggregation approach, a DL algorithm that correctly identifies which features are the most important for the network's predictions in an unstructured tabular data set | Curated Microarray Database (CuMiDa) [113] | Tabular unstructured data | DL | LSTM |
| Chereda et al. [114] | 2021 | Extend the procedure of LRP to make it available for Graph-CNN (GCN) and test its applicability on a large breast cancer data set | Gene Expression Omnibus (GEO) [115] | Genomics data | DL | Graph-CNN |

## 5.2.2 | Gradient-weighted class activation mapping (Grad-CAM)

Grad-CAM [122] is a local backpropagation-based method that uses the feature maps produced by the last layer of a CNN to create a coarse localization class-specific heatmap where the hot part corresponds to a particular class. Grad-CAM is based on CAM but is not limited to fully connected CNNs. Grad-CAM can be applied to any CNN architecture without retraining or architectural modification as long as the layers are differentiable.

As detailed in Table 7, (4/30) studies used Grad-CAM to determine how accurately the CNN model localized the breast masses. The authors of article [104] have also investigated the performance of the model-agnostic LIME in conjunction with Grad-CAM to investigate the aspects and utilities of two different XAI methods in explaining the misclassification of breast masses. The results highlight the usability of XAI in understanding the mechanism of used AI models and their failures, which can provide valuable insights toward explainable CAD systems. In Gerbasi et al. [107], the authors also implemented Deep SHAP, a high-speed approximation algorithm for computing SHAP values in DL models, to produce maps visually interpreting the classification results, which in the maps, pink pixels strongly contributed to the final predicted class (malignant), and the blue pixels contributed to the prediction of opposite class (benign).

## 5.2.3 | Gradient-weighted class activation mapping++ (Grad-CAM++)

Grad-CAM++ [123] is a local backpropagation-based method built upon Grad-CAM to enhance visual explanations of CNN. Compared to Grad-CAM, it provides better visual explanations of model predictions in terms of better localization of objects and explaining occurrences of multiple objects of a class in a single image [123]. As listed in Table 8, only one study (1/30) used Grad-CAM++. To et al. [110] developed an ensemble learning-based approach to locate cancerous regions in DUV whole-slide images (WSI). It used Grad-CAM++ on a pretrained DenseNet169 model to generate regional significance maps to classify each WSI confidently as cancerous or benign.

### 5.2.4 | Layer-wise relevance propagation (LRP)

LRP [124] is a local propagation-based approach. LRP calculates the relevance score for a specific output at the classifier layer. It proceeds backward, exploits the DL structure, and calculates each neuron's explanatory factors (relevance R) for each layer during the backward pass until it reaches the input image [18, 124]. Based on the computed relevance score, LRP generates a heatmap with highlighted critical regions that can be used to explain the prediction. Two studies (2/30) have implemented LRP; the details are described in Table 9.

Grisci et al. [112] introduced relevance aggregation, an XAI approach based on LRP that combines the relevance derived from several samples as learned by a neural network and generates scores for each input feature. The study results showed that the model could correctly identify which input features or relevant ones are the most important for the model's predictions, facilitate knowledge discovery, and help identify incorrect or irrelevant rules or machine biases in the case of the poorly trained implemented AI model. Chereda et al. [114] extended the procedure of LRP to make it available for GCN to explain its decisions. They tested it on a large BC genomic data set. They showed that the model, named graph layer-wise relevance propagation (GLRP), provides patient-specific molecular subnetworks that agree with clinical knowledge and can identify common, novel, and potentially druggable drivers of tumor progression.

In summary, 12 out of 30 studies used model-specific XAI methods, and CAM and Grad-CAM were the most used models, respectively.

### 5.3 | Clinical applications

Figure 6 illustrates the diverse applications of each XAI method across various clinical scenarios. The studies primarily focused on either diagnosing/classifying breast cancer or conducting survival/prognosis analyses of patients. Within these study types, image recognition techniques were employed on radiology data, or alternative approaches utilizing clinical and demographic data were explored. Notably, SHAP was frequently utilized in clinical data analysis studies rather than image recognition studies. This preference may be attributed to the computational resource intensity of SHAP, posing challenges in handling the high-dimensional feature space inherent in image data. Conversely, techniques such as CAM and Grad-CAM are computationally less intensive, which makes them a better choice for image processing tasks, especially in real-time applications. In diagnosis/classification studies, the primary objective was to employ supervised learning methods for distinguishing between healthy and diseased patients, facilitating early detection. XAI models played a crucial role in helping clinicians comprehend and validate intricate patterns and features that influence diagnostic outcomes. In survival/prognosis models, clinicians sought to predict the onset of events such as mortality or metastasis in patients. XAI methods proved instrumental in interpreting and elucidating the contribution of each factor to a patient's outcome measure. This interpretability makes the models more understandable, usable, and trustworthy for both clinicians and patients, fostering a perception and interpretation of the predictions and building confidence in the decision-making process.

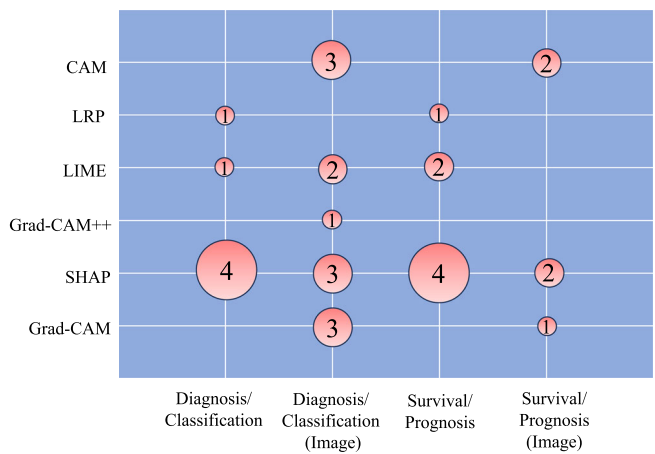

**FIGURE 6** Explainable artificial intelligence (XAI) methods are applied in different clinical applications of breast cancer literature. Numbers inside the bubbles represent the number of studies.

### 5.4 | Future directions

The rapid evolution of AI models, as evidenced by advanced frameworks such as GPT and generative AI-based models, is significantly transforming the healthcare applications landscape. As these models continue to advance and become more intricate, the necessity for XAI methods becomes increasingly imperative. In the healthcare domain, where precision and interpretability are of paramount importance, the demand for robust XAI techniques is expected to grow. Future research should prioritize the refinement and advancement of XAI methodologies to effectively uncover the intricacies of advanced AI models in healthcare contexts. The synergy between the rapid advancements in AI technologies and the evolving landscape of XAI is crucial, shaping the trajectory of personalized healthcare and ensuring that these innovative models translate into tangible benefits for both clinicians and patients.

## 5.5 │ Study limitations

This systematic scoping review has some limitations that warrant consideration. While efforts were made to minimize publication bias, excluding non-English language articles, non-access, and gray literature may have resulted in the omission of some valuable information. Additionally, despite our best efforts to construct a comprehensive search strategy across multiple databases using combinations of Boolean search strings and MeSH terms, the diverse terminology associated with XAI methods and breast cancer may have led to the inadvertent omission of certain studies. Moreover, we only investigated the existing established XAI methods; however, XAI schemes based on or independent of these methods can be observed in a few studies. To ensure the integrity and credibility of the study, we did not consider some of the studies with low or no citations in this survey.

## 6 │ CONCLUSIONS

We systematically reviewed breast cancer studies that successfully implemented the existing XAI methods to their model predictors. In summary, SHAP was the most used model-agnostic method. The frequent use of this method with tree-based ensemble ML models is related to the speed and compatibility that SHAP provided for these models. Grad-CAM and CAM were widely used model-specific XAI methods in these studies. We noticed that other explanatory methods, as provided in Appendix A, have not been used in breast cancer studies and can still be examined and compared as future endeavors.

Additionally, the XAI methods used in the selected studies only provided a sanity check to the model's predictor results. As was mentioned in the introduction, finding the biases in the model and data can be achieved using explainability methods that were either missing or only mentioned in a few of the studies and should be investigated for further studies. Moreover, although the clinical applications of XAI methods were investigated in our study, the results generated by these methods were not evaluated by oncologists. Therefore, to provide trustworthiness, the reliability of the results through clinical evaluation is needed. Researchers have already used XAI domain-specific explanations to improve understanding, interpretation, trustworthiness, and reliability of the results in different medical domains for evaluating health interventions [125], disease causal pathway analysis [126], mental health surveillance and precision resource allocation [127], precision dermatology and disease diagnosis [128], immune response predictors [129], and investigating the links between socioenvironmental risk factors and Alzheimer's disease [130].

Potential challenges associated with the application of XAI, especially when dealing with complex multimodal clinical and medical data, include but are not limited to the availability of data in appropriate temporal and geographic resolutions; its representativeness, diversity, and types of modalities involved, semantic heterogeneity, fusion of heterogeneous data streams, AI-readiness of clinical data sets [131], and algorithmic and human biases in explanations that addressing them can increase the efficiency and acceptance of multimodal XAI schemes.

Addressing these challenges is key to the widespread acceptance of multimodal XAI models and algorithms in cancer care delivery and treatment.

## AUTHOR CONTRIBUTIONS

**Amirehsan Ghasemi**: Conceptualization (lead); data curation (lead); formal analysis (lead); investigation (lead); methodology (lead); validation (lead); visualization (lead); writing—original draft (lead); writing—review and editing (lead). **Soheil Hashtarkhani**: Conceptualization (supporting); data curation (supporting); formal analysis (supporting); investigation (supporting); methodology (supporting); validation (supporting); visualization (supporting); writing—original draft (supporting); writing—review and editing (supporting). **David L. Schwartz**: Conceptualization (supporting); funding acquisition (supporting); investigation (supporting); resources (supporting); supervision (supporting); writing—original draft (supporting); writing—review and editing (supporting). **Arash Shaban-Nejad**: Conceptualization (lead); funding acquisition (lead); methodology (supporting); project administration (lead); resources (lead); supervision (lead); writing—original draft (supporting); writing—review and editing (supporting).

## ACKNOWLEDGMENTS

This study was supported by the Center for Biomedical Informatics at the University of Tennessee Health Science Center.

## CONFLICT OF INTEREST STATEMENT

The authors declare no conflict of interest.

## DATA AVAILABILITY STATEMENT

The authors declare that the data supporting the findings of this study are available within the paper.

## ETHICS STATEMENT

Not applicable.

## INFORMED CONSENT

Not applicable.

## ORCID

*Amirehsan Ghasemi*  http://orcid.org/0000-0001-9288-6731

*Soheil Hashtarkhani*  https://orcid.org/0000-0001-7750-6294

*David L. Schwartz*  https://orcid.org/0000-0002-7235-5586

*Arash Shaban-Nejad*  http://orcid.org/0000-0003-2047-4759

## REFERENCES

1. Zhang B, Vakanski A, Xian M. BI-RADS-NET-V2: a composite multi-task neural network for computer-aided diagnosis of breast cancer in ultrasound images with semantic and quantitative explanations. IEEE Access. 2023;11:79480–94. https://doi.org/10.1109/ACCESS.2023.3298569

2. Hayes-Roth F. Rule-based systems. Commun ACM. 1985;28(9):921–32. https://doi.org/10.1145/4284.4286

3. Howard A, Zhang C, Horvitz E. Addressing bias in machine learning algorithms: a pilot study on emotion recognition for intelligent systems. In: 2017 IEEE workshop on advanced robotics and its social impacts (ARSO). Austin, TX, USA: IEEE, 2017 Workshop on Advanced Robotics and its Social Impacts (ARSO); 2017. p. 1–7.

4. van Giffen B, Herhausen D, Fahse T. Overcoming the pitfalls and perils of algorithms: a classification of machine learning biases and mitigation methods. J Bus Res. 2022;144:93–106. https://doi.org/10.1016/j.jbusres.2022.01.076

5. Shaban-Nejad A, Michalowski M, Brownstein JS, Buckeridge DL. Guest editorial explainable AI: towards fairness, accountability, transparency and trust in healthcare. IEEE J Biomed Health Inform. 2021;25(7):2374–5. https://doi.org/10.1109/JBHI.2021.3088832

6. Quinn TP, Senadeera M, Jacobs S, Coghlan S, Le V. Trust and medical AI: the challenges we face and the expertise needed to overcome them. J Am Med Inform Asso. 2020;28(4):890–4. https://doi.org/10.1093/jamia/ocaa268

7. Pan SJ, Yang Q. A survey on transfer learning. IEEE Trans Knowl Data Eng. 2010;22(10):1345–59. https://doi.org/10.1109/TKDE.2009.191

8. He K, Zhang X, Ren S, Sun J. Identity mappings in deep residual networks. In: Leibe B, Matas J, Sebe N, Welling M, editors. Computer vision—ECCV 2016. Cham: Springer International Publishing; 2016. p. 630–45.

9. Mishkin D, Matas J. All you need is a good init. arXiv preprint arXiv:151106422. 2015.

10. Ioffe S, Szegedy C. Batch normalization: accelerating deep network training by reducing internal covariate shift. In: Bach F, Blei D, editors. Proceedings of the 32nd international conference on machine learning vol. 37 of Proceedings of Machine Learning Research. Lille: PMLR; 2015. p. 448–56. Available from: https://proceedings.mlr.press/v37/ioffe15.html

11. Yu D, Xiong W, Droppo J, Stolcke A, Ye G, Li J, et al. Deep convolutional neural networks with layer-wise context expansion and attention. In: Interspeech. San Francisco, California, USA: Proc. Interspeech; 2016. p. 17–21. https://doi.org/10.21437/Interspeech.2016-251

12. Gunning D, Aha D. DARPA's explainable artificial intelligence (XAI) program. AI Magazine. 2019;40(2):44–58. https://doi.org/10.1609/aimag.v40i2.2850

13. Shaban-Nejad A, Michalowski M, Buckeridge DL. Explainability and interpretability: Keys to deep medicine. In: Shaban-Nejad A, Michalowski M, Buckeridge DL, editors. Explainable AI in healthcare and medicine: building a culture of transparency and accountability. Cham: Springer International Publishing; 2021. p. 1–10.

14. Shaban-Nejad A, Michalowski M, Buckeridge D. Explainable AI in healthcare and medicine: building a culture of transparency and accountability. Stud Comp Intel. 2020;914: 344. https://doi.org/10.1007/978-3-030-53352-6

15. Luo W, Phung D, Tran T, Gupta S, Rana S, Karmakar C, et al. Guidelines for developing and reporting machine learning predictive models in biomedical research: a multidisciplinary view. J Med Internet Res. 2016;18(12):e323. https://doi.org/10.2196/jmir.5870

16. Miller T. Explanation in artificial intelligence: insights from the social sciences. Artif Intel. 2019;267:1–38. https://doi.org/10.1016/j.artint.2018.07.007

17. Bhattacharya A. Applied machine learning explainability techniques: make ML models explainable and trustworthy for practical applications using LIME, SHAP, and more. Packt Publishing Ltd; 2022. Available from: https://download.packt.com/free-ebook/9781803246154

18. Loh HW, Ooi CP, Seoni S, Barua PD, Molinari F, Acharya UR. Application of explainable artificial intelligence for healthcare: a systematic review of the last decade (2011–2022). Comp Methods Prog Biomed. 2022;226:107161. https://doi.org/10.1016/j.cmpb.2022.107161

19. Bharati S, Mondal MRH, Podder P. A review on explainable artificial intelligence for healthcare: why, how, and when? IEEE Trans Artif Intel. 2023;5:1–15. https://doi.org/10.1109/TAI.2023.3266418

20. Di Martino F, Delmastro F. Explainable AI for clinical and remote health applications: a survey on tabular and time series data. Artif Intel Rev. 2023;56(6):5261–315. https://doi.org/10.1007/s10462-022-10304-3

21. Hauser K, Kurz A, Haggenmüller S, Maron RC, von Kalle C, Utikal JS, et al. Explainable artificial intelligence in skin cancer recognition: a systematic review. Eur J Cancer. 2022;167:54–69. https://doi.org/10.1016/j.ejca.2022.02.025

22. James G, Witten D, Hastie T, Tibshirani R, et al. An introduction to statistical learning. vol. 112. New York, NY, USA: Springer; 2013.

23. Geurts P, Ernst D, Wehenkel L. Extremely randomized trees. Mach Learn. 2006;63:3–42. https://doi.org/10.1007/s10994-006-6226-1

24. Breiman L. Random forests. Mach Learn. 2001;45:5–32. https://doi.org/10.1023/A:1010933404324

25. Chen T, Guestrin C. XGBoost: a scalable tree boosting system. In: Proceedings of the 22nd ACM SIGKDD international conference on knowledge discovery and data mining. KDD '16. New York: Association for Computing Machinery; 2016. p. 785–94.

26. Ke G, Meng Q, Finley T, Wang T, Chen W, Ma W, et al. LightGBM: a highly efficient gradient boosting decision tree. In: Guyon I, Luxburg UV, Bengio S, Wallach H, Fergus R, Vishwanathan S, et al., editors. Advances in neural information processing systems. vol. 30. Long Beach, CA, USA: Curran Associates, Inc.; 2017. p. 3149–57. Available from: https://proceedings.neurips.cc/paper_files/paper/2017/file/6449f44a102fde848669bdd9eb6b76fa-Paper.pdf

27. Freund Y, Schapire RE. A decision-theoretic generalization of on-line learning and an application to boosting. J Comp Syst Sci. 1997;55(1):119–39. https://doi.org/10.1006/jcss.1997.1504

28. Dorogush AV, Ershov V, Gulin A. CatBoost: gradient boosting with categorical features support. arXiv preprint arXiv:181011363. 2018.

29. Sagi O, Rokach L. Ensemble learning: a survey. WIREs Data Min Knowl Disc. 2018;8(4):e1249. https://doi.org/10.1002/widm.1249

30. Shaban-Nejad A, Michalowski M, Bianco S. Multimodal artificial intelligence: next wave of innovation in healthcare and medicine. Stud Comp Intel. 2022;1060:1–9.

31. Cybenko GV. Approximation by superpositions of a sigmoidal function. Math Control Signals Syst. 1989;2:303–14. https://doi.org/10.1007/BF02551274

32. Hornik K, Stinchcombe M, White H. Multilayer feedforward networks are universal approximators. Neural Netw. 1989;2(5):359–66. https://doi.org/10.1016/0893-6080(89)90020-8

33. Baydin AG, Pearlmutter BA, Radul AA, Siskind JM. Automatic differentiation in machine learning: a survey. J Mach Learn Res. 2018;18:1–43. https://doi.org/10.48550/arXiv.1502.05767

34. Lecun Y, Bottou L, Bengio Y, Haffner P. Gradient-based learning applied to document recognition. Proc IEEE. 1998;86(11):2278–324. https://doi.org/10.1109/5.726791

35. Simonyan K, Zisserman A. Very deep convolutional networks for large-scale image recognition. arXiv preprint arXiv:14091556. 2014.

36. Krizhevsky A, Sutskever I, Hinton GE. ImageNet classification with deep convolutional neural networks. Commun ACM. 2017;60(6):84–90. https://doi.org/10.1145/3065386

37. Chollet F. Xception: deep learning with depthwise separable convolutions. In: 2017 IEEE conference on computer vision and pattern recognition (CVPR). Honolulu, HI, USA: IEEE; 2017. p. 1800–7. https://doi.org/10.1109/CVPR.2017.195

38. Szegedy C, Liu W, Jia Y, Sermanet P, Reed S, Anguelov D, et al. Going deeper with convolutions. In: Proceedings of the IEEE conference on computer vision and pattern recognition (CVPR). Boston, MA, USA: IEEE; 2015. p. 1–9. https://doi.org/10.1109/CVPR.2015.7298594

39. Szegedy C, Vanhoucke V, Ioffe S, Shlens J, Wojna Z. Rethinking the inception architecture for computer vision. In: Proceedings of the IEEE conference on computer vision and pattern recognition (CVPR). Las Vegas, NV, USA: IEEE; 2016. p. 2818–26. https://doi.org/10.1109/CVPR.2016.308

40. Szegedy C, Ioffe S, Vanhoucke V, Alemi A. Inception-v4, inception-ResNet and the impact of residual connections on learning. Proc AAAI Conf Artif Intel. 2017;31(1):4278–84. https://doi.org/10.1609/aaai.v31i1.11231

41. He K, Zhang X, Ren S, Sun J. Deep residual learning for image recognition. In: 2016 IEEE conference on computer vision and pattern recognition (CVPR). Las Vegas, NV, USA: IEEE; 2016. p. 770–8. https://doi.org/10.1109/CVPR.2016.90

42. Xie S, Girshick R, Dollar P, Tu Z, He K. Aggregated residual transformations for deep neural networks. In: Proceedings of the IEEE conference on computer vision and pattern recognition (CVPR). Honolulu, HI, USA: IEEE; 2017. p. 5987–95. https://doi.org/10.1109/CVPR.2017.634

43. Zhang H, Wu C, Zhang Z, Zhu Y, Lin H, Zhang Z, et al. ResNeSt: split-attention networks. In: Proceedings of the IEEE/CVF conference on computer vision and pattern recognition (CVPR) workshops; 2022. p. 2736–46.

44. Ronneberger O, Fischer P, Brox T. U-Net: Convolutional networks for biomedical image segmentation. In: Navab N, Hornegger J, Wells WM, Frangi AF, editors. Medical image computing and computer-assisted intervention–MICCAI 2015. Cham: Springer International Publishing; 2015. p. 234–41.

45. Kipf TN, Welling M. Semi-supervised classification with graph convolutional networks. arXiv. https://doi.org/10.48550/arXiv.1609.02907

46. Huang G, Liu Z, Pleiss G, Weinberger KQ. Convolutional networks with dense connectivity. IEEE Trans Pattern Anal Mach Intel. 2022;44(12):8704–16. https://doi.org/10.1109/TPAMI.2019.2918284

47. Tan M, Le Q. EfficientNet: rethinking model scaling for convolutional neural networks. In: Chaudhuri K, Salakhutdinov R, editors. Proceedings of the 36th international conference on machine learning vol. 97 of proceedings of machine learning research. Long Beach, CA, USA: PMLR; 2019. p. 6105–14. Available from: https://proceedings.mlr.press/v97/tan19a.html

48. Howard AG, Zhu M, Chen B, Kalenichenko D, Wang W, Weyand T, et al. Mobilenets: efficient convolutional neural networks for mobile vision applications. arXiv preprint arXiv:170404861. 2017.

49. Sandler M, Howard A, Zhu M, Zhmoginov A, Chen LC. MobileNetV2: inverted residuals and linear bottlenecks. In: Proceedings of the IEEE conference on computer vision and pattern recognition (CVPR). Salt Lake City, UT, USA: IEEE; 2018. p. 4510–20. https://doi.org/10.1109/CVPR.2018.00474

50. Howard A, Sandler M, Chu G, Chen LC, Chen B, Tan M, et al. Searching for MobileNetV3. In: Proceedings of the IEEE/CVF international conference on computer vision (ICCV) Seoul, South Korea: IEEE; 2019. p. 1314–24. https://doi.org/10.1109/ICCV.2019.00140

51. Zhang X, Zhou X, Lin M, Sun J. ShuffleNet: an extremely efficient convolutional neural network for mobile devices. In: Proceedings of the IEEE conference on computer vision and pattern recognition (CVPR); 2018.

52. Iandola FN, Han S, Moskewicz MW, Ashraf K, Dally WJ, Keutzer K. SqueezeNet: AlexNet-level accuracy with 50x fewer parameters and <0.5 MB model size. arXiv:160207360. 2016.

53. Hochreiter S, Schmidhuber J. Long short-term memory. Neural Comput. 1997;9(8):1735–80. https://doi.org/10.1162/neco.1997.9.8.1735

54. Schuster M, Paliwal KK. Bidirectional recurrent neural networks. IEEE Trans Signal Proces. 1997;45(11):2673–81. https://doi.org/10.1109/78.650093

55. Cho K, van Merriënboer B, Gulcehre C, Bahdanau D, Bougares F, Schwenk H, et al. Learning phrase representations using RNN encoder–decoder for statistical machine translation. In: Moschitti A, Pang B, Daelemans W, editors. Proceedings of the 2014 conference on empirical methods in natural language processing (EMNLP). Doha, Qatar:

Association for Computational Linguistics; 2014. p. 1724–34. Available from: https://aclanthology.org/D14-1179

56. van der Veer SN, Riste L, Cheraghi-Sohi S, Phipps DL, Tully MP, Bozentko K, et al. Trading off accuracy and explainability in AI decision-making: findings from 2 citizens' juries. J Am Med Inform Assoc. 2021;28(10):2128–38. https://doi.org/10.1093/jamia/ocab127

57. Pintelas E, Livieris IE, Pintelas P. A grey-box ensemble model exploiting black-box accuracy and white-box intrinsic interpretability. Algorithms. 2020;13(1):1–17. https://doi.org/10.3390/a13010017

58. Bennetot A, Franchi G, Ser JD, Chatila R, Díaz-Rodríguez N. Greybox XAI: a neural-symbolic learning framework to produce interpretable predictions for image classification. Knowl Based Syst. 2022;258:109947. https://doi.org/10.1016/j.knosys.2022.109947

59. Wanner J, Herm LV, Heinrich K, Janiesch C, Zschech P. White, grey, black: effects of XAI augmentation on the confidence in ai-based decision support systems. Forty-First International Conference on Information Systems, India; 2020.

60. Bohlin TP. Practical grey-box process identification: theory and applications. London, UK: Springer Science & Business Media; 2006.

61. Vilone G, Longo L. Classification of explainable artificial intelligence methods through their output formats. Mach Learn Knowl Extract. 2021;3(3):615–61. https://doi.org/10.3390/make3030032

62. Adadi A, Berrada M. Peeking inside the black-box: a survey on explainable artificial intelligence (XAI). IEEE Access. 2018;6:52138–60. https://doi.org/10.1109/ACCESS.2018.2870052

63. Samek W, Montavon G, Lapuschkin S, Anders CJ, Müller KR. Explaining deep neural networks and beyond: a review of methods and applications. Proc IEEE. 2021;109(3):247–78. https://doi.org/10.1109/JPROC.2021.3060483

64. Clement T, Kemmerzell N, Abdelaal M, Amberg M. XAIR: a systematic metareview of explainable AI (XAI) aligned to the software development process. Mach Learn Knowl Extract. 2023;5(1):78–108. https://doi.org/10.3390/make5010006

65. Molnar C. Interpretable machine learning. 2nd ed.; 2022 [cited 2024 Jan 15]. Available from: https://christophm.github.io/interpretable-ml-book

66. Zilke JR, Loza Mencía E, Janssen F. DeepRED—rule extraction from deep neural networks. In: Calders T, Ceci M, Malerba D, editors. Discovery science. Cham: Springer International Publishing; 2016. p. 457–73.

67. Craven M, Shavlik J. Extracting tree-structured representations of trained networks. In: Touretzky D, Mozer MC, Hasselmo M, editors. Advances in neural information processing systems. vol. 8. MIT Press; 1995. Available from: https://proceedings.neurips.cc/paper_files/paper/1995/file/45f31d16b1058d586fc3be7207b58053-Paper.pdf

68. Werbos PJ. Beyond regression: new tools for prediction and analysis in the behavioral sciences. PhD thesis, Committee on Applied Mathematics, Harvard University, Cambridge, MA. 1974.

69. Rumelhart DE, Hinton GE, Williams RJ. Learning internal representations by error propagation. San Diego, CA, USA: California University of San Diego La Jolla Institute for Cognitive Science; 1985.

70. LeCun Y. A learning scheme for asymmetric threshold networks. Proc Cognit. 1985;85(537):599–604.

71. Parker DB. Learning-logic: casting the cortex of the human brain in silicon. Tech Rep. 1985;47.

72. Moher D, Liberati A, Tetzlaff J, Altman DG. The PRISMA Group. Preferred reporting items for systematic reviews and meta-analyses: the PRISMA statement. Ann Int Med. 2009;151(4):264–9. https://doi.org/10.7326/0003-4819-151-4-200908180-00135

73. Tober M. PubMed, ScienceDirect, Scopus or Google Scholar—which is the best search engine for an effective literature research in laser medicine? Med Laser Appl. 2011;26(3):139–44. https://doi.org/10.1016/j.mla.2011.05.006

74. Chakraborty D, Ivan C, Amero P, Khan M, Rodriguez-Aguayo C, Başağaoğlu H, et al. Explainable artificial intelligence reveals novel insight into tumor microenvironment conditions linked with better prognosis in patients with breast cancer. Cancers. 2021;13(14):1–15. https://doi.org/10.3390/cancers13143450

75. Moncada-Torres A, van Maaren MC, Hendriks MP, Siesling S, Geleijnse G. Explainable machine learning can outperform cox regression predictions and provide insights in breast cancer survival. Sci Rep. 2021;11. https://doi.org/10.1038/s41598-021-86327-7

76. Rezazadeh A, Jafarian Y, Kord A. Explainable ensemble machine learning for breast cancer diagnosis based on ultrasound image texture features. Forecasting. 2022;4(1):262–74. https://doi.org/10.3390/forecast4010015

77. Al-Dhabyani W, Gomaa M, Khaled H, Fahmy A. Dataset of breast ultrasound images. Data Brief. 2020;28:104863. https://doi.org/10.1016/j.dib.2019.104863

78. Nahid AA, Raihan MJ, Bulbul AAM. Breast cancer classification along with feature prioritization using machine learning algorithms. Health Technol. 2022;12(6):1061–9. https://doi.org/10.1007/s12553-022-00710-6

79. Yu H, Chen F, Lam KO, Yang L, Wang Y, Jin JY, et al. Potential determinants for radiation-induced lymphopenia in patients with breast cancer using interpretable machine learning approach. Front Immunol. 2022;13. https://doi.org/10.3389/fimmu.2022.768811

80. Meshoul S, Batouche A, Shaiba H, AlBinali S. Explainable multi-class classification based on integrative feature selection for breast cancer subtyping. Mathematics. 2022;10(22). https://doi.org/10.3390/math10224271

81. Kumar S, Das A. Peripheral blood mononuclear cell derived biomarker detection using explainable artificial intelligence (XAI) provides better diagnosis of breast cancer. Comp Biol Chem. 2023;104:107867. https://doi.org/10.1016/j.compbiolchem.2023.107867

82. Silva-Aravena F, NúñezDelafuente H, Gutiérrez-Bahamondes JH, Morales J. A hybrid algorithm of ML and XAI to prevent breast cancer: a strategy to support decision making. Cancers. 2023;15(9):1–18. https://doi.org/10.3390/cancers15092443

83. Nindrea RD, Usman E, Katar Y, Darma IY, Warsiti, Hendriyani H, et al. Dataset of Indonesian women's reproductive, high-fat diet and body mass index risk factors for breast cancer. Data in Brief. 2021;36:107107. https://doi.org/10.1016/j.dib.2021.107107

84. Massafra R, Fanizzi A, Amoroso N, Bove S, Comes MC, Pomarico D, et al. Analyzing breast cancer invasive disease event classification through explainable artificial intelligence.

2770183, 2024, 1, Downloaded from https://onlinelibrary.wiley.com/doi/10.1002/cai2.136 by University Of Tennessee Health Science Center, Wiley Online Library on [12/07/2024]. See the Terms and Conditions (https://onlinelibrary.wiley.com/terms-and-conditions) on Wiley Online Library for rules of use; OA articles are governed by the applicable Creative Commons License

Front Med. 2023;10. https://doi.org/10.3389/fmed.2023.1116354

85. Vrdoljak J, Boban Z, Baric D, Segvic D, Kumric M, Avirovic M, et al. Applying explainable machine learning models for detection of breast cancer lymph node metastasis in patients eligible for neoadjuvant treatment. Cancers. 2023;15(3). https://doi.org/10.3390/cancers15030634

86. Mohi Uddin KM, Biswas N, Rikta ST, Dey SK, Qazi A. XML-LightGBMDroid: a self-driven interactive mobile application utilizing explainable machine learning for breast cancer diagnosis. Eng Rep. 2023;5:e12666. https://doi.org/10.1002/eng2.12666

87. Wolberg MOSN William, Street W. Breast Cancer Wisconsin (Diagnostic). UCI Machine Learning Repository. 1995 [cited 2024 Jan 15]. Available from: https://doi.org/10.24432/C5DW2B.

88. Zhao X, Jiang C. The prediction of distant metastasis risk for male breast cancer patients based on an interpretable machine learning model. BMC Med Informat Decision Making. 2023;23(1):74. https://doi.org/10.1186/s12911-023-02166-8

89. Cordova C, Muñoz R, Olivares R, Minonzio JG, Lozano C, Gonzalez P, et al. HER2 classification in breast cancer cells: a new explainable machine learning application for immuno-histochemistry. Oncol Lett. 2023;25(2):1–9. https://doi.org/10.3892/ol.2022.13630

90. Kaplun D, Krasichkov A, Chetyrbok P, Oleinikov N, Garg A, Pannu HS. Cancer cell profiling using image moments and neural networks with model agnostic explainability: a case study of breast cancer histopathological (BreakHis) database. Mathematics. 2021;9(20):1–20. https://doi.org/10.3390/math9202616

91. Spanhol FA, Oliveira LS, Petitjean C, Heutte L. A dataset for breast cancer histopathological image classification. IEEE Trans Biomed Eng. 2016;63(7):1455–62. https://doi.org/10.1109/TBME.2015.2496264

92. Saarela M, Jauhiainen S. Comparison of feature importance measures as explanations for classification models. SN Appl Sci. 2021;3. https://doi.org/10.1007/s42452-021-04148-9

93. Adnan N, Zand M, Huang THM, Ruan J. Construction and evaluation of robust interpretation models for breast cancer metastasis prediction. IEEE/ACM Trans Comp Biol Bioinform. 2022;19(3):1344–53. https://doi.org/10.1109/TCBB.2021.3120673

94. Staiger C, Cadot S, Györffy B, Wessels L, Klau G. Current composite-feature classification methods do not outperform simple single-genes classifiers in breast cancer prognosis. Front Genet. 2013;4. https://doi.org/10.3389/fgene.2013.00289

95. Maouche I, Terrissa LS, Benmohammed K, Zerhouni N. An explainable AI approach for breast cancer metastasis prediction based on clinicopathological data. IEEE Trans Biomed Eng. 2023;70:1–9. https://doi.org/10.1109/TBME.2023.3282840

96. Slaoui M, Mouh FZ, Ghanname I, Razine R, El Mzibri M, Amrani M. Outcome of breast cancer in Moroccan young women correlated to clinic-pathological features, risk factors and treatment: a comparative study of 716 cases in a single institution. PLoS One. 2016;11(10):1–14. https://doi.org/10.1371/journal.pone.0164841

97. Deshmukh S, Behera BK, Mulay P, Ahmed EA, Al-Kuwari S, Tiwari P, et al. Explainable quantum clustering method to model medical data. Knowl Based Syst. 2023;267:110413. https://doi.org/10.1016/j.knosys.2023.110413

98. Qi X, Zhang L, Chen Y, Pi Y, Chen Y, Lv Q, et al. Automated diagnosis of breast ultrasonography images using deep neural networks. Med Image Anal. 2019;52:185–98. https://doi.org/10.1016/j.media.2018.12.006

99. Zhou LQ, Wu XL, Huang SY, Wu GG, Ye HR, Wei Q, et al. Lymph node metastasis prediction from primary breast cancer US images using deep learning. Radiology. 2020;294(1):19–28. https://doi.org/10.1148/radiol.2019190372

100. Huang Z, Zhu X, Ding M, Zhang X. Medical image classification using a light-weighted hybrid neural network based on PCANet and DenseNet. IEEE Access. 2020;8:24697–712. https://doi.org/10.1109/ACCESS.2020.2971225

101. Xi P, Guan H, Shu C, Borgeat L, Goubran R. An integrated approach for medical abnormality detection using deep patch convolutional neural networks. Vis Comput. 2020;36(9):1869–82. https://doi.org/10.1007/s00371-019-01775-7

102. Kim J, Kim HJ, Kim C, Lee JH, Kim KW, Park YM, et al. Weakly-supervised deep learning for ultrasound diagnosis of breast cancer. Sci Rep. 2021;11:24382. https://doi.org/10.1038/s41598-021-03806-7

103. El Adoui M, Drisis S, Benjelloun M. Multi-input deep learning architecture for predicting breast tumor response to chemotherapy using quantitative MRI images. Int J Comp Assist Radiol Surg. 2020;15. https://doi.org/10.1007/s11548-020-02209-9

104. Hussain SM, Buongiorno D, Altini N, Berloco F, Prencipe B, Moschetta M, et al. Shape-based breast lesion classification using digital tomosynthesis images: the role of explainable artificial intelligence. Appl Sci. 2022;12(12). https://doi.org/10.3390/app12126230

105. Bevilacqua V, Brunetti A, Guerriero A, Trotta GF, Telegrafo M, Moschetta M. A performance comparison between shallow and deeper neural networks supervised classification of tomosynthesis breast lesions images. Cogn Syst Res. 2019;53:3–19. https://doi.org/10.1016/j.cogsys.2018.04.011

106. Agbley BLY, Li JP, Haq AU, Bankas EK, Mawuli CB, Ahmad S, et al. Federated fusion of magnified histopathological images for breast tumor classification in the Internet of medical things. IEEE J Biomed Health Inform. 2023:1–12. https://doi.org/10.1109/JBHI.2023.3256974

107. Gerbasi A, Clementi G, Corsi F, Albasini S, Malovini A, Quaglini S, et al. DeepMiCa: automatic segmentation and classification of breast MIcroCAlcifications from mammograms. Comp Methods Prog Biomed. 2023;235:107483. https://doi.org/10.1016/j.cmpb.2023.107483

108. Moreira IC, Amaral I, Domingues I, Cardoso A, Cardoso MJ, Cardoso JS. INbreast: toward a full-field digital mammographic database. Acad Radiol. 2012;19(2):236–48. https://doi.org/10.1016/j.acra.2011.09.014

109. Lee RP, Markantonakis K, Akram RN. Provisioning software with hardware-software binding. In: Proceedings of the 12th international conference on availability, reliability and security. ARES '17. New York, NY, USA: Association for Computing Machinery; 2017. Available from: https://doi.org/10.1145/3098954.3103158

110. To T, Lu T, Jorns JM, Patton M, Schmidt TG, Yen T, et al. Deep learning classification of deep ultraviolet fluorescence images toward intra-operative margin assessment in breast

cancer. Front Oncol. 2023;13. https://doi.org/10.3389/fonc.2023.1179025

111. Lu T, Jorns JM, Patton M, Fisher R, Emmrich A, Doehring T, et al. Rapid assessment of breast tumor margins using deep ultraviolet fluorescence scanning microscopy. J Biomed Opt. 2020;25(12):126501. https://doi.org/10.1117/1.JBO.25.12.126501

112. Grisci BI, Krause MJ, Dorn M. Relevance aggregation for neural networks interpretability and knowledge discovery on tabular data. Inform Sci. 2021;559:111–29. https://doi.org/10.1016/j.ins.2021.01.052

113. Feltes BC, Chandelier EB, Grisci BI, Dorn M. CuMiDa: an extensively curated microarray database for benchmarking and testing of machine learning approaches in cancer research. J Comp Biol. 2019;26(4):376–86. https://doi.org/10.1089/cmb.2018.0238

114. Chereda H, Bleckmann A, Menck K, Perera-Bel J, Stegmaier P, Auer F, et al. Explaining decisions of graph convolutional neural networks: patient-specific molecular subnetworks responsible for metastasis prediction in breast cancer. Genome Med. 2021;13. https://doi.org/10.1186/s13073-021-00845-7

115. Barrett T, Wilhite SE, Ledoux P, Evangelista C, Kim IF, Tomashevsky M, et al. NCBI GEO: archive for functional genomics data sets—update. Nucl Acids Res. 2012;41(D1): D991–5. https://doi.org/10.1093/nar/gks1193

116. Lundberg SM, Lee SI. A unified approach to interpreting model predictions. In: Guyon I, Luxburg UV, Bengio S, Wallach H, Fergus R, Vishwanathan S, et al., editors. Advances in neural information processing systems. vol. 30. Curran Associates, Inc.; 2017. p. 1–10. Available from: https://proceedings.neurips.cc/paper_files/paper/2017/file/8a20a8621978632d76c43dfd28b67767-Paper.pdf

117. Shapley LS. A value for n-person games. Princeton: Princeton University Press; 1953. p. 307–18. Available from: https://doi.org/10.1515/9781400881970-018

118. Winter E. Chapter 53 The shapley value vol. 3 of handbook of game theory with economic applications. Elsevier; 2002. p. 2025–54. Available from: https://www.sciencedirect.com/science/article/pii/S1574000502030163

119. Ribeiro MT, Singh S, Guestrin C. "Why should I trust you?": explaining the predictions of any classifier. In: DeNero J, Finlayson M, Reddy S, editors. Proceedings of the 22nd ACM SIGKDD international conference on knowledge discovery and data mining. KDD '16. New York, NY, USA: Association for Computing Machinery; 2016. p. 1135–44. Available from: https://doi.org/10.1145/2939672.2939778

120. Sadeghi Z, Alizadehsani R, Cifci M, Kausar S, Rehman R, Mahanta P, et al. A brief review of explainable artificial intelligence in healthcare. Comput Electr Eng. 2024;118. https://doi.org/10.1016/j.compeleceng.2024.109370

121. Zhou B, Khosla A, Lapedriza A, Oliva A, Torralba A. Learning deep features for discriminative localization. In: 2016 IEEE conference on computer vision and pattern recognition (CVPR). Las Vegas, NV, USA: IEEE; 2016. p. 2921–9.

122. Selvaraju RR, Cogswell M, Das A, Vedantam R, Parikh D, Batra D. Grad-CAM: visual explanations from deep networks via gradient-based localization. Int J Comput Vision.

2020;128(2):336–59. https://doi.org/10.1007/s11263-019-01228-7

123. Chattopadhay A, Sarkar A, Howlader P, Balasubramanian VN. Grad-CAM++: generalized gradient-based visual explanations for deep convolutional networks. In: 2018 IEEE winter conference on applications of computer vision (WACV). Lake Tahoe, NV, USA: IEEE; 2018. p. 839–47.

124. Bach S, Binder A, Montavon G, Klauschen F, Müller KR, Samek W. On pixel-wise explanations for non-linear classifier decisions by layer-wise relevance propagation. PLoS One. 2015;10(7):1–46. https://doi.org/10.1371/journal.pone.0130140

125. Brenas JH, Shaban-Nejad A. Health intervention evaluation using semantic explainability and causal reasoning. IEEE Access. 2020;8:9942–52. https://doi.org/10.1109/ACCESS.2020.2964802

126. Brakefield WS, Ammar N, Shaban-Nejad A. An urban population health observatory for disease causal pathway analysis and decision support: underlying explainable artificial intelligence model. JMIR Form Res. 2022;6(7):e36055. https://doi.org/10.2196/36055

127. Ammar N, Shaban-Nejad A. Explainable artificial intelligence recommendation system by leveraging the semantics of adverse childhood experiences: proof-of-concept prototype development. JMIR Med Inform. 2020;8(11):e18752. https://doi.org/10.2196/18752

128. Chanda T, Hauser K, Hobelsberger S, Bucher TC, Garcia CN, Wies C, et al. Dermatologist-like explainable AI enhances trust and confidence in diagnosing melanoma. Nat Commun. 2024;15(1):524. https://doi.org/10.1038/s41467-023-43095-4

129. Borole P, Rajan A. Building trust in deep learning-based immune response predictors with interpretable explanations. Commun Biol. 2024;7(1):279. https://doi.org/10.1038/s42003-024-05968-2

130. Fania A, Monaco A, Amoroso N, Bellantuono L, Cazzolla Gatti R, Firza N, et al. Machine learning and XAI approaches highlight the strong connection between $O_3$ and $NO_2$ pollutants and Alzheimer's disease. Sci Rep. 2024;14(1): 5385. https://doi.org/10.1038/s41598-024-55439-1

131. Ng MY, Youssef A, Miner AS, Sarellano D, Long J, Larson DB, et al. Perceptions of data set experts on important characteristics of health data sets ready for machine learning: a qualitative study. JAMA Netw Open. 2023;6(12):e2345892. https://doi.org/10.1001/jamanetworkopen.2023.45892

132. Ribeiro MT, Singh S, Guestrin C. Anchors: high-precision model-agnostic explanations. In: AAAI conference on artificial intelligence. New Orleans, LA, USA: AAAI; 2018. p. 1527–35. https://doi.org/10.1609/aaai.v32i1.11491

133. Zeiler MD, Fergus R. Visualizing and understanding convolutional networks. In: Fleet D, Pajdla T, Schiele B, Tuytelaars T, editors. Computer vision—ECCV 2014. Cham: Springer International Publishing; 2014. p. 818–33.

134. Friedman JH. Greedy function approximation: a gradient boosting machine. Ann Statist. 2001;29(5):1189–232. https://doi.org/10.1214/aos/1013203451

135. Wachter S, Mittelstadt B, Russell C. Counterfactual explanations without opening the black box: automated decisions and the GDPR. Harv JL Tech. 2017;31:841. https://doi.org/10.48550/arXiv.1711.00399

136. Buitinck L, Louppe G, Blondel M, Pedregosa F, Mueller A, Grisel O, et al. API design for machine learning software: experiences from the scikit-learn project. In: ECML PKDD workshop: languages for data mining and machine learning; 2013. p. 108–22. https://doi.org/10.48550/arXiv.1309.0238

137. Klaise J, Looveren AV, Vacanti G, Coca A. Alibi explain: algorithms for explaining machine learning models. J Mach Learn Res. 2021;22(181):1–7. Available from: https://jmlr.org/papers/v22/21-0017.html

138. Sundararajan M, Taly A, Yan Q. Axiomatic attribution for deep networks. In: Precup D, Teh YW, editor. Proceedings of the 34th international conference on machine learning. vol. 70 of proceedings of machine learning research. PMLR; 2017. p. 3319–28. Available from: https://proceedings.mlr.press/v70/sundararajan17a.html

139. Shrikumar A, Greenside P, Kundaje A. Learning important features through propagating activation differences. ICML'17: Proceedings of the 34th International Conference on Machine Learning. Vo. 70. 2017. p. 3145–53.

140. Montavon G, Lapuschkin S, Binder A, Samek W, Müller KR. Explaining nonlinear classification decisions with deep Taylor decomposition. Pattern Recogn. 2017;65:211–22. https://doi.org/10.1016/j.patcog.2016.11.008

141. Springenberg JT, Dosovitskiy A, Brox T, Riedmiller M. Striving for simplicity: the all convolutional net. arXiv. 2015.

142. Erhan D, Bengio Y, Courville A, Vincent P. Visualizing higher-layer features of a deep network. Univ Montreal. 2009;1341(3):1.

143. Kim B, Wattenberg M, Gilmer J, Cai CJ, Wexler J, Viégas FB, et al. Interpretability beyond feature attribution: quantitative testing with concept activation vectors (TCAV). Proceedings of the 35th International Conference on Machine Learning (PMLR 80, 2018), Stockholm, Sweden; 2017.

144. Huang Q, Yamada M, Tian Y, Singh D, Chang Y. GraphLIME: local interpretable model explanations for graph neural networks. IEEE Trans Knowl Data Eng. 2023;35(7):6968–72. https://doi.org/10.1109/TKDE.2022.3187455

**How to cite this article:** Ghasemi A, Hashtarkhani S, Schwartz DL, Shaban-Nejad A. Explainable artificial intelligence in breast cancer detection and risk prediction: a systematic scoping review. Cancer Innov. 2024;3:e136. https://doi.org/10.1002/cai.136

## APPENDIX A: POPULAR MODEL-AGNOSTIC AND MODEL-SPECIFIC XAI METHODS

The summary of model-agnostic and model-specific XAI methods is listed as follows.

See Tables A1–A4.

**TABLE A1**  Model-agnostic explainable artificial intelligence (XAI) methods.

| Model-agnostic method | Acronym | Type of XAI | Scope of XAI | Technique |
|---|---|---|---|---|
| SHapley Additive exPlanations [116] | SHAP | Feature importance | Local, Global | Game-theory |
| Local interpretable model agnostic explanations [119] | LIME | Feature importance | Local | Surrogate model |
| Anchors [132] | N/A | Feature importance | Local | Surrogate model |
| Occlusion sensitivity [133] | Occlusion | Feature importance | Local | Perturbation-based |
| Partial dependence plots [134] | PDP | Visual explanations | Global | Marginalization |
| Counterfactuals [135] | N/A | Example-based XAI | Global | Data point |
| Rule extraction [66] | N/A | White-Box model | Global | White-box |
| Tree extraction [67] | N/A | White-Box model | Global | White-box |

**TABLE A2**  Model-agnostic: Code/Toolbox.

| Model-agnostic method | Code/Toolbox |
|---|---|
| SHAP | https://github.com/slundberg/shap |
| LIME | https://github.com/marcotcr |
| Anchors | https://github.com/marcotcr/anchor |
| Occlusion | Can be found here: https://www.mathworks.com |
| PDP | Can be found here [136] |
| Counterfactuals | Can be found here [137]→https://github.com/SeldonIO/alibi |

**TABLE A3** Model-specific explainable artificial intelligence (XAI) methods.

| Model-specific method | Acronym | Black-box model | Type of XAI | Scope of XAI | Technique |
|---|---|---|---|---|---|
| Class activation map [121] | CAM | Convolutional neural network (CNN) | Feature importance | Local | Propagation-based |
| Gradient-weighted class activation mapping [122] | Grad-CAM | CNN | Feature importance | Local | Propagation-based |
| Gradient-weighted class activation mapping++ [123] | Grad-CAM++ | CNN | Feature importance | Local | Propagation-based |
| Integrated gradients [138] | IG | All DL models | Feature importance | Local | Propagation-based |
| Deep Learning Important FeaTures [139] | DeepLIFT | All DL models | Feature importance | Local | Propagation-based |
| Layerwise relevance propagation [124] | LRP | All DL models | Feature importance | Local | Propagation-based |
| Deep Taylor decomposition [140] | DTD | All DL models | Feature importance | Local | Propagation-based |
| Guided backpropagation [141] | GBP | All DL models | Feature importance | Local | Propagation-based |
| Activation maximization [142] | N/A | All DL models | Feature importance | Global | Propagation-based |
| Testing with concept activation vectors [143] | TCAV | All DL models | Feature importance | Global | Concept-based |
| Model explanation for graph neural networks [144] | GraphLIME | GNN | Feature importance | Local | Surrogate model |

**TABLE A4** Model-specific explainable artificial intelligence methods: Code/Toolbox.

| Model-specific method | Code/Toolbox |
|---|---|
| CAM | https://github.com/zhoubolei/CAM |
| Grad-CAM | https://github.com/ramprs/grad-cam/ |
| Grad-CAM++ | https://github.com/adityac94/Grad_CAM_plus_plus |
| IG | https://github.com/ankurtaly/Integrated-Gradients |
| DeepLIFT | https://github.com/kundajelab/deeplift |
| LRP, DTD | https://github.com/chr5tphr/zennit<br>https://github.com/albermax/innvestigate |
| Guided backpropagation | https://github.com/mateuszbuda/ALL-CNN |
| TCAV | https://github.com/tensor?ow/tcav |
| GraphLIME | https://github.com/WilliamCCHuang/GraphLIME |